\documentstyle[12pt, psfig]{article}

\newcommand{\NP}{{\em Nucl.\ Phys.\ }}
\newcommand{\PL}{{\em Phys.\ Lett.\ }}

\newcommand{\PRP}{{\em Phys.\ Rep.\ }}

\newcommand{\PRL}{{\em Phys.\ Rev.\ Lett.\ }}
\newcommand{\IJMP}{{\em Int.\ J.\ Mod.\ Phys.\ }}

\newcommand{\tr}{{\rm Tr}}

\newcommand{\fslash}{F\!\!\!\!/\ }
\newcommand{\inn}{\!\cdot\!}

\begin{document}
\pagestyle{plain}
\setcounter{page}{1}

\baselineskip16pt

\begin{titlepage}

\begin{flushright}
PUPT-1612\\
hep-th/9604065
\end{flushright}
\vspace{20 mm}

\begin{center}
{\huge Decay of Excited D-branes}

\vspace{5mm}

\end{center}

\vspace{10 mm}

\begin{center}
{\large A.~Hashimoto and I.R.~Klebanov}

\vspace{3mm}

Joseph Henry Laboratories\\
Princeton University\\
Princeton, New Jersey 08544

\end{center}

\vspace{2cm}

\begin{center}
{\large Abstract}
\end{center}
We calculate the leading order interactions of massless D-brane
excitations. Their 4-point functions are found to be identical
to those found in type I theory. The amplitude for two massless
D-brane fluctuations to produce a massless closed string is found to
possess interesting new structure. As a function of its single kinematic
invariant, it displays an infinite sequence of alternating zeros and poles.
At high transverse momenta, this amplitude decays exponentially,
indicating a growing effective thickness of the D-brane.
This amplitude is the leading process by which non-extrtemal D-branes
produce Hawking radiation.

\noindent

\vspace{2cm}
\begin{flushleft}
April 1996
\end{flushleft}
\end{titlepage}
\newpage

\renewcommand{\baselinestretch}{1.1}  

\section{Introduction}
\label{Intro}

Non-perturbative duality symmetries of string theory
\cite{filq,sen,schwarz,Schwarz2,duff,chpt,witten,strominger,jhas}
require that there exist $p$-brane soliton solutions carrying
Ramond-Ramond $(p+1)$-form charge.  Solitons of this kind were found
some time ago in the effective low-energy description of superstrings
provided by supergravity \cite{dh,mdjl,chs,hs}.  Quite recently, it
was realized by Polchinski \cite{polchinski} that the exact stringy
description of the R-R charged $p$-branes is provided by the Dirichlet
branes \cite{dlp,joeReview}. The essence of the D-brane construction
is that the perturbative formulation of type II strings in the
background of a $p$-brane soliton requires an introduction of world
sheets with boundaries. The $p+1$ world volume coordinates are
supplied with the conventional Neumann boundary conditions, while the
remaining $9-p$ coordinates have the Dirichlet boundary conditions,
$X^i = X^i_0$, where $X^i_0$ is the position of the center of the
soliton. The remarkable simplicity of this description of string
solitons has led to rapid progress in non-perturbative string theory
\cite{witinst,doug,vafa,bsv,horw,as,schmid}.

One advantage of the D-brane description is that scattering of closed
strings off $p$-brane solitons may now be treated perturbatively.
Leading order calculations of $1\rightarrow 1$ scattering amplitudes
for massless closed string states were first carried out in
\cite{KT,GHKM}. Explicit formulae were obtained by integrating two-point
functions of bulk operators on a disk. They exhibit a typical stringy
structure: a delicate interplay between infinite sequences of
$t$-channel and $s$-channel poles. The $s$-channel poles occur in the
physical region (for real momenta) and correspond to excited
D-branes. The high-energy behavior of the scattering amplitudes
exhibits the well-known stringy features: the Regge behavior of
forward scattering and the exponential fall-off of fixed angle
scattering \cite{regge,kks}. These features continue to hold for
perturbative corrections to the $1\rightarrow 1$ amplitudes
\cite{Barbon}.

A systematic approach to all massless two-point functions on a disk
was recently developed in \cite{gm} where it was shown that there is a
direct relation between four-point functions of type I theory and two-point
functions of type II theory in a D-brane background.  Since the former
amplitudes are well-known for all massless states, a simple
transformation gives the massless two-point functions with most
general polarizations. The results are in agreement with the direct
calculations in \cite{KT,GHKM}.

The purpose of this note is to calculate two more types of D-brane
amplitudes that are simply related to the massless four-point functions
of the type I theory. The first type is quite obvious: it is the
four-point function for the massless internal modes of Dirichlet
$p$-branes.  Such modes are present for $p>0$ and come in two classes:
the world volume gauge fields and the transverse excitations. Since
the two are related by T-duality, their four-point functions may be simply
obtained from the type I formula. Indeed, both the D-brane and type I
amplitudes are calculated on a disk with four boundary operators.

A somewhat less obvious example are the D-brane disk amplitudes with
one bulk and two boundary operators.  This provides the leading order
amplitude for either the absorption of a graviton by
a D-brane or the decay of an excited D-brane into a massless closed
string state and the unexcited D-brane \cite{DasMathur96,cm}.  This decay is
possible whenever the two massless waves on the D-brane run in
different directions.\footnote{If the massless
waves run in the same direction, then we are dealing with a BPS state
that does not decay.}  The analysis of this process, which is the
leading manifestation of the Hawking radiation in the D-brane picture,
is of obvious interest.  We find that the kinematics of this process
is characterized by a single invariant,
$$t=-(p_1+p_2)^2$$
where $p_1$ and $p_2$ are the momenta of the open string states. Thus,
$t$ is the length-squared of the total momentum parallel to the
$p$-brane.  The general formula 
for the three-point function is\footnote{We omit the normalization factor 
which is proportional to the string coupling constant.}
\begin{equation}
A = {\Gamma (-2 t)\over \Gamma^2 (1-t)} K(1, 2, 3)
\end{equation}
which for large $t$ decays as
\begin{equation}
A = 2^{-2t} t^{-3/2} K(1,2,3).
\end{equation}
The polarization dependence $K$ is obtained from that of type I
four-point function by a simple transformation. The fact that $A$
falls off exponentially at large $t$ is another indication of the
smearing of D-branes at high energies (the expanding string halo
surrounding a D-brane). A related but nevertheless remarkable fact, is
that a highly energetic excited state is less likely to decay than a
low energy excitation. We will comment on the possible significance of
this in the discussion.

In section 2 we show how the four-point functions for the D-brane
internal modes are read off from the type I four-point functions. In
section 3 we analyze the three-point functions of one bulk and two
boundary operators on a disk.

\section{World volume scattering of massless D-brane excitations}

In the $0$ superghost picture,
the vertex operator describing a world volume photon
of momentum $k$ is

\begin{equation}
V_0^A = (\partial_t X^A +
i k\inn \psi \psi^A) e^{i k\cdot X} (t)
\label{photop}\end{equation}
where the index $A$ spans the world volume directions,
$A=0, \ldots, p$. The momentum $k$ is, of course,
also restricted to lie within the world volume directions.
The vertex operator for a scalar describing a transverse
$p$-brane oscillation is related to (\ref{photop}) by T-duality,

\begin{equation}
V_0^i = (\partial_n X^i +
i k\inn \psi \psi^i) e^{i k\cdot X} (t)
\label{scalop}\end{equation}
where $i=p+1, \ldots, 9$.

While the operators (\ref{photop}), (\ref{scalop})
are inserted on the boundary of a half-plane, it is often convenient
to double the half-plane to the entire complex plane \cite{KLS88}.
In this formalism, all operators are built out of the
purely holomorphic field. The advantage of this for our purposes is that
the photon and scalar operators may be assembled into
the following form,

\begin{eqnarray}
V_{-1}^\mu(z,2k) &=& 
e^{-\phi} \psi^\mu e^{i\, 2k \cdot X} (z) \nonumber \\
V_0^\mu(z,2k) &=& (\partial X^\mu +
i\, 2k\inn\psi\, \psi^\mu) e^{i\, 2k \cdot X} (z) 
\label{ops}\end{eqnarray}
where the index $\mu$ now takes all possible values
from $0$ to $9$ (we have explicitly 
written down both the $-1$ and $0$ superghost picture operators). 
This
is not surprising since the world volume scalars and the gauge fields
arise as dimensional reduction of $d=10$ super Yang-Mills theory.  
The four-point function of D-brane
oscillations assumes the form
\begin{eqnarray}
\lefteqn{A(\zeta_1, k_1; \zeta_2, k_2; \zeta_3, k_3; \zeta_4, k_4)} \nonumber \\
&& \sim
\int \frac{dx_1 dx_2 dx_3 dx_4}{V_{CKG}}
\langle \zeta_1\inn V_{0}(2k_1,x_1)\,
\zeta_2\inn V_{0}(2k_2,x_2) \,
\zeta_3\inn V_{-1}(2k_3,x_3) \,
\zeta_4\inn V_{-1}(2k_4,x_4) \rangle
\end{eqnarray}
where $V_{CKG}$ is the volume of the $SL(2,R)$ conformal Killing
group. Contractions are to be evaluated with the Green functions,
\footnote{We work in units where $\alpha'$=2 for
both type I and type II strings.}
\begin{eqnarray}
\langle X^\mu(z) X^\nu(w)\rangle &=&- \eta^{\mu\nu} \ln (z-w) \nonumber \\
\langle \psi^\mu(z) \psi^\nu(w)\rangle &=&- {\eta^{\mu\nu}
\over z-w }\nonumber \\
\langle \phi (z) \phi (w)\rangle &=&- \ln (z-w) 
\end{eqnarray}
This expression for $A$ is identical to the four-photon correlation
function of type I theory!  As a result, all correlation functions
involving both the world volume gauge fields and the scalar
fluctuations of D-branes may be read off from the type I massless NS
four-point function \cite{SchwarzPhysRep},
\begin{equation}
A(\zeta_1, k_1; \zeta_2, k_2; \zeta_3, k_3; \zeta_4, k_4)=
{\Gamma (4 k_1\inn k_2) \Gamma (4 k_1\inn k_4)\over
\Gamma (1+ 4 k_1\inn k_2+ 4 k_1\inn k_4) }
K (\zeta_1, k_1; \zeta_2, k_2; \zeta_3, k_3; \zeta_4, k_4)
\label{typeone}
\end{equation}
where
\begin{eqnarray}
\lefteqn{K = 4 k_2\inn k_3\,  k_2\inn k_4\, \zeta_1\inn\zeta_2\,
\zeta_3\inn\zeta_4} \nonumber \\
&&\ \ \ 4 k_1\inn k_2 (\zeta_1\inn k_4 \zeta_3\inn k_2
\zeta_2\inn\zeta_4 + 
\zeta_2\inn k_3 \zeta_4\inn k_1 \zeta_1\inn\zeta_3
+\zeta_1\inn k_3 \zeta_4\inn k_2 \zeta_2\inn\zeta_3
+\zeta_2\inn k_4 \zeta_3\inn k_1 \zeta_1\inn\zeta_4)\nonumber \\
&&\ \ \ \ \ + \{(1234) \rightarrow (1324)\} + \{(1234) \rightarrow (1432)\}
\label{kinematic.factor}
\end{eqnarray}

In order to apply this formula to D-branes we restrict all momenta
$k_l$ to lie in the world volume directions. For scattering of four
world volume photons, we also restrict all $\zeta_l$ to lie in the
world volume directions. The resulting amplitude is obviously given by
(\ref{typeone}).  If, on the other hand, we are interested in the
scattering of scalars, then we choose the corresponding polarizations
$\zeta$ to lie in the transverse directions. The four-scalar amplitude
simplifies because $\zeta_l\inn k_m =0$. The polarization dependence
is
\begin{equation}
K =
-4\,k_2\inn k_3\, k_2\inn k_4\, \zeta_1\inn\zeta_2\, \zeta_3\inn\zeta_4
-4\,k_2\inn k_3\, k_3\inn k_4\, \zeta_1\inn\zeta_3\, \zeta_2\inn\zeta_4
-4\,k_4\inn k_3 k_2\inn k_4\, \zeta_1\inn\zeta_4\, \zeta_2\inn\zeta_3\, 
\end{equation}
The non-trivial structure of the world volume $2\rightarrow 2$
scattering indicates that the internal dynamics of D-branes is highly
non-trivial.  The $s$ and $t$ channel poles are due to massive states
of the open strings attached to D-branes, which are typically excluded
from the low-energy D-brane effective actions \cite{ck}.  This
suggests that behind the simplicity of the Dirac-Born-Infeld effective
actions for D-branes \cite{dlp,schmid,Tseytlin96} a more complex dynamics is
hidden.

The only remaining non-vanishing NS amplitude is for two gauge fields
and two scalars. It is a simple matter to extract it from
(\ref{typeone}) by restricting polarizations appropriately.
Correlation functions involving the Ramond (fermion) operators on the
boundary may be discussed along similar lines.

\section{Hawking radiation from excited D-branes}

In this section, we compute the amplitude for the collision of two
massless open strings stuck to a D-brane to produce a massless
closed string state in the bulk.  To begin, let us restrict to
the NS sector and assign momenta $p_1$ and $p_2$ to the open string
states and momentum $q$ for the closed string. The D-brane kinematics
is such that only the momentum along the direction parallel to the
brane is conserved,
\begin{equation}
p_1 + p_2 + q_\parallel = 0 \ .
\end{equation}
The open string momenta $p_1$ and $p_2$ are 
restricted to lie within the D-brane world volume.
Ordinarily in massless three-point amplitudes, conservation of momentum
constrains the kinematics completely, but in the presence of D-branes,
the non-conservation of momenta in the directions transverse to the brane
gives rise to some freedom.  Here, from conservation of longitudinal
momentum, it follows that there is exactly one kinematic invariant in
this problem, which we call $t$,
\begin{equation}
t = 2 p_1 \inn q = 2 p_2 \inn q = - 2 p_1 \inn p_2 \ .
\end{equation}

The leading order contribution to this amplitude is 
evaluated on a disk
with two operators on the boundary and one in the bulk.  We proceed by
mapping the disk 
to upper half-plane. The amplitude may be written
generally in the form
\begin{equation}
A = \int \frac{d z_1\, d z_2\, d^2 z_3}{V_{CKG}} \langle 
V_1(z_1)
V_2(z_2)
V_3(z_3,\bar{z}_3)
\rangle \label{gen.amp}
\end{equation}
where $z_1$ and $z_2$ are integrated only along the real axis. $V_1$
and $V_2$ are the boundary
vertex operators given in (\ref{photop}), (\ref{scalop}),
while $V_3$ is the is the usual massless closed string vertex operator,
\begin{eqnarray}
V_1(z_1) &=&  \xi^1_\mu V_0^\mu(z_1,p_1) \\
V_2(z_2) &=&  \xi^2_\nu V_0^\nu(z_2,p_2) \\
V_3(z_3,\bar{z}_3) & =& \varepsilon_{\sigma\lambda}\, 
             V_{-1}^\sigma(z_3,q)\,
             \tilde V_{-1}^\lambda(\bar{z}_3,q)
\end{eqnarray}
where $\xi$ and $\varepsilon$ are polarization tensors. Following the
approach of \cite{KT,GHKM,gm} we extend the 
calculation to the entire complex plane. This is accomplished by the 
following substitutions for the antiholomorphic fields
in terms of the holomorphic fields,
\begin{equation}
\tilde X^\mu (\bar z)\rightarrow D^\mu_\nu X^\nu (\bar z)\ ,\quad
\tilde \psi^\mu (\bar z)\rightarrow D^\mu_\nu \psi^\nu (\bar z)\ ,\quad
\tilde \phi (\bar z)\rightarrow \phi (\bar z)\ .
\end{equation}
$D^\mu_\nu$, as defined in \cite{gm}, is the diagonal matrix
whose first $p+1$ entries, corresponding to the longitudinal directions,
are equal to $1$, while the remaining $9-p$ entries, corresponding to
the transverse directions, are equal to $-1$.
The amplitude (\ref{gen.amp}) becomes
\begin{equation}
A = \int \frac{dz_1\, dz_2\, dz_3\, d\bar{z}_3}{V_{CKG}}\,
\xi_\mu^1 \xi_\nu^2 \varepsilon_{\sigma \lambda} {D^\lambda}_\eta
\langle V_0^\mu(z_1,2p_1) 
        V_0^\nu(z_2,2p_2) 
        V_{-1}^\sigma(z_3,q) 
        V_{-1}^\eta(\bar{z}_3, D\inn q) \rangle
\label{integral.expression}
\end{equation}
The integrand of the above expression is 
brought to the standard form
\begin{equation}
\langle 
\zeta_1 \inn V_0(z_1, 2k_1)\ 
\zeta_2 \inn V_0(z_2, 2k_2)\ 
\zeta_3 \inn V_{-1}(z_3, 2k_3)\
\zeta_4 \inn V_{-1}(z_4, 2k_4) \rangle
\end{equation}
by the following identifications,
\begin{equation}
\begin{array}{llll}
2 p_1     \rightarrow 2 k_1 &
2 p_2     \rightarrow 2 k_2 & 
q         \rightarrow 2 k_3 & 
D \inn q \rightarrow 2 k_4\\
\xi_1  \rightarrow \zeta_1 &
\xi_2  \rightarrow \zeta_2 &
\multicolumn{2}{l}{\varepsilon \inn  D  \rightarrow \zeta_3 \otimes \zeta_4}  \\
\end{array}.
\label{nsns.identify}
\end{equation}
The correlation functions of this form have appeared before in
computations of open string four-point functions and closed string
two-point functions in the D-brane background. 
The only new features of our 
calculation are the kinematics
and a different prescription for fixing the residual $SL(2,R)$ on
the world sheet.

For future use, it is convenient to introduce Mandelstam variables
\begin{equation}
s = 4k_1 \inn k_2 = 4 k_3 \inn k_4, \qquad
t = 4k_1 \inn k_3 = 4 k_2 \inn k_4, \qquad
u = 4k_1 \inn k_4 = 4 k_2 \inn k_3.
\end{equation}
In previous work (boundary 4-point function
or D-brane 2-point function), 
any pair of them constituted an independent set of kinematic
invariants.  Note that $t = 4 k_1 \inn k_3 = 2 p_1 \inn q$ is the same
kinematic quantity as introduced earlier.

In the calculation of the closed string two-point function, for
example, it was convenient to fix the operators at $\{z_1, z_2, z_3,
z_4\} = \{iy, -iy, i,-i\}$.  When this is done and the appropriate
Jacobian factors are inserted, (\ref{integral.expression}) becomes
\begin{equation}
A = \int dy 
\left[ \frac{4y}{(1+y)^2} \right]^s
\left[ \frac{(1-y)^2}{(1+y)^2} \right]^t 
\left[ \frac{1}{1-y^2} a_1 - \frac{1-y}{4 y (1+y)} a_2 \right]
\label{y.integral}
\end{equation}
where $a_1$ and $a_2$ contain the dependences on polarization and
momenta, but are independent of $y$.  Performing the $y$-integral, one
finds
\begin{equation}
A= \frac{\Gamma(t) \Gamma(s)}{\Gamma(1+s+t)}(s a_1 - t a_2).
\end{equation}
It was shown in \cite{gm} that $(s a_1 - t a_2)$ is identical to the
kinematic factor (\ref{kinematic.factor}) from open string four-point
amplitudes.

For the current problem of three-point function for two open strings and
one closed string, it is convenient to fix the operators at $\{z_1,
z_2, z_3, z_4\} = \{x, -x, i,-i\}$, which corresponds to fixing the
closed string vertex at $z=i$ and constraining the locations of 
the open string
vertex operators on the real axis.  Notice that this corresponds
simply to a change of variables, $y = ix$.  Also, recalling that there
is only one kinematic invariant in this problem, we should set $s = 4
k_1 \inn k_2 = -2 t$.  After these replacements
(\ref{y.integral}) takes the from:
\begin{equation}
A=\int_{-\infty}^{\infty} dx \left[ \frac{(1 + x^2)^2}{16 x^2} \right]^t
\frac{1}{1+x^2} \left(a_1 + \frac{a_2}{2}\right)
\end{equation}
where only the terms even under $x \rightarrow -x$ are kept. The integral
is again doable and we arrive at 
\begin{equation}
A =  (-2 t  a_1 - t a_2)  \frac{\Gamma[-2t]}{\Gamma[1-t]^2}
\end{equation}
It is clear that $(-2t a_1 - t a_2)$ is the same kinematic factor $K =
(s a_1 - t a_2)$ we encountered above with $s = -2 t$.  So what we
found is
\begin{equation}
A =  \frac{\Gamma[-2t]}{\Gamma[1-t]^2}
 K(1, 2, 3)
 \label{main.result}
\end{equation}
This is the main result of this section.  
The kinematic factor is obtained
from that of type I theory, (\ref{kinematic.factor}), by the
identifications (\ref{nsns.identify}).  
The explicit formula for the Neveu-Schwarz amplitudes is, therefore,
\begin{eqnarray}
K(1, 2, 3) & = &
  \left[\rule{0ex}{2ex} t\,\left( - q\inn \xi_2\,\xi_1\inn  D \inn q\,
         g^{\mu \nu}   - 
      q\inn \xi_1\,\xi_2\inn  D\inn q\,g^{\mu\nu} - 
      4\,\xi_1\inn \xi_2\,p_1^\nu \,p_2^\mu  - 
      4\,\xi_1 \inn \xi_2\,p_1^\mu \,p_2^\nu  \right.  \right. \nonumber \\  
&&\qquad- 
      2\,p_1 \inn \xi_2\,q^\nu \,\xi_1^\mu  + 
      4\,q\inn \xi_2\,p_1^\nu \,\xi_1^\mu  - 
      2\,p_1\inn \xi_2\, (D \inn q)^\mu \,\xi_1^\nu + 
      4\,\xi_2\inn  D\inn q\,p_1^\mu \,\xi_1^\nu  \nonumber \\
&&\qquad \left. - 
      2\,p_2\inn \xi_1\,q^\nu \,\xi_2^\mu  + 
      4\,q\inn \xi_1\,p_2^\nu \,\xi_2^\mu  - 
      2\,p_2\inn \xi_1\,(D \inn q)^\mu \,\xi_2^\nu  + 
      4\,\xi_1 \inn  D \inn q\,p_2^\mu \,\xi_2^\nu  \right) \nonumber \\
&&\ \ \   +  \left.
   {t^2}\,\left( -4\,\xi_1\inn \xi_2\,g^{\mu \nu} + 
      8\,\xi_1^\nu \,\xi_2^\mu  + 8\,\xi_1^\mu \,\xi_2^\nu
       \right) \rule{0ex}{2ex} \right] (\varepsilon \inn D)_{\mu \nu}
\end{eqnarray}
Gauge invariance follows automatically from the
structure of type I kinematic factor.  When we take the polarization
of the closed string to be strictly transverse to the D-brane, the
kinematic factor simplifies drastically and one is left with
\begin{equation}
A \sim \frac{\Gamma[-2t]}{\Gamma[1-t]^2} t^2
\left(\xi^1 \inn  \varepsilon  \inn \xi^2 +
\xi^2 \inn  \varepsilon  \inn \xi^1 \right)  
\end{equation}

A few comments are in order about states in the Ramond sector.  In
\cite{gm}, it was noted that the type I four-point amplitude and the
closed string two-point amplitude are {\em identical} under suitable
identification of polarizations and momenta. Here, we have found that
the three-point function for two open and one closed strings are
related by simple change of variables $y = ix$, or equivalently,
change in the integration contour of the modular parameter of the
amplitude.  This change of variables leads to a different
structure of the universal beta-function factor as we saw above, but
does not affect the polarization dependence of these amplitudes.
Therefore, just as in \cite{gm}, one should expect to find appropriate
kinematic factors from the type I theory to appear in the
three-point amplitude for states in the Ramond sector with the identification
of polarization and momenta very similar to what was discussed in
\cite{gm}. Let us illustrate this with some examples.

First, consider two Neveu-Schwarz open strings bosons colliding to
become a closed string Ramond-Ramond $n$-form $F_{\mu_1 \mu_2 \ldots
\mu_n}$.  In this case, the correlation function of interest
will be of the form
\begin{equation}
\xi_\mu \xi_\nu (P\fslash M)^{AB}
\langle 
 V_0^\mu(z_1, 2p_1)\,
 V_{-1}^\nu(z_2, 2p_2)\,
 V_{-1/2\, A}(z_3,q)\, 
 V_{-1/2\, B}(\bar{z}_3,D \inn q)
\rangle
\end{equation}
where 
\begin{eqnarray}
\fslash & =& \frac{1}{n!} 
F_{\mu_1 \mu_2 \cdots \mu_n} 
\gamma^{\mu_1} \gamma^{\mu_2} \cdots \gamma^{\mu_n}, \\
M & = & \gamma^0 \gamma^1 \cdots \gamma^p,
\end{eqnarray}
and $P = (1 + \gamma_{11})/2$ projects the spinors onto definite
chirality. The amplitude of interest is found to be
\begin{equation}
A =  \frac{\Gamma[-2t]}{\Gamma[1-t]^2}K(1, 2, 3)
\end{equation}
The kinematic factor $K$ is obtained from that in
type I theory, \cite{SchwarzPhysRep}
\begin{eqnarray}
\lefteqn{K(u_1,\,\zeta_2,\,\zeta_3,\,u_4)=
k_1\inn k_4\,\bar{u}_1\gamma\inn \zeta_2\gamma\inn(k_3+k_4)\gamma\inn\zeta_3 u_4}
\nonumber \\
&&-2\,k_1\inn k_2\,\left(\bar{u}_1\gamma\inn\zeta_3u_4\,k_3\inn\zeta_2
-\bar{u}_1\gamma\inn\zeta_2u_4\,k_2\inn\zeta_3-\bar{u}_1\gamma\inn k_3u_4\,\zeta_2\inn\zeta_3\right)
\end{eqnarray}
by the following substitutions
\begin{equation}
\begin{array}{llll}
2 k_2     \rightarrow 2 p_1 &
2 k_3     \rightarrow 2 p_2 & 
2 k_4          \rightarrow q & 
2 k_1     \rightarrow D \inn q \\
\zeta_2   \rightarrow \xi_1  &
\zeta_3   \rightarrow \xi_2  &
\multicolumn{2}{l}{u_4 \otimes u_1\rightarrow  P \fslash M } \\
\end{array}
\end{equation}
Thus, we find the following explicit formula,
\begin{eqnarray}
K(1, 2, 3)  &=&
t\,
\tr\left[P \fslash M 
\left( \rule{0ex}{2ex} \gamma\inn \xi_1\gamma\inn(p_2+q/2)\gamma\inn\xi_2
+ 4( p_2\inn\xi_1 ) \gamma\inn\xi_2  \right. \right. \nonumber \\
&&\qquad \left. \left. \rule{0ex}{2ex} - 4(p_1\inn\xi_2) \gamma\inn\xi_1
- 4 (\xi_1\inn\xi_2) \gamma\inn p_2 \right) \right]
\end{eqnarray}

Similarly, consider two Ramond open string fermions with polarizations
$v_1^A$ and $v_2^B$ colliding to become a NS-NS boson with
polarization $\varepsilon_{\mu \nu}$. Now, the correlation function
takes the form
\begin{equation}
v_1^A v_2^B (\varepsilon \inn D)_{\mu \nu}
\langle 
V_{-1/2\, A}(z_1, 2p_1)\, 
V_{-1/2\, B}(z_2, 2p_2)\,
V_0^\mu(z_3, q)\,
V_{-1}^\nu(\bar{z}_3, D \inn q)
\rangle
\end{equation}
This can be gotten from the same kinematic factor as above, but with
identifications
\begin{equation}
\begin{array}{llll}
2 p_1     \rightarrow 2 k_4 &
2 p_2     \rightarrow 2 k_1 & 
q         \rightarrow 2 k_2 & 
D \inn q \rightarrow 2 k_3\\
v_1 \rightarrow u_4 &
v_2 \rightarrow u_1 &
\multicolumn{2}{l}{(\varepsilon \inn D) \rightarrow \zeta_2  \otimes  \zeta_3} \\
\end{array}
\end{equation}
Thus, the amplitude for two open string fermions to produce a NS-NS closed
string state is
\begin{eqnarray}
A & \sim & \frac{\Gamma[-2t]}{\Gamma[1-t]^2}\, t\, \left[\rule{0ex}{2ex}
 \bar{v}_2\gamma^\mu \gamma\inn(D \inn q+ 2p_1)\gamma^\nu v_1
 + 4 \bar{v}_2\gamma^\nu v_1\,(D \inn q)^\mu \right. \nonumber \\
&& \qquad  \left. \rule{0ex}{2ex}
- 4\bar{v}_2\gamma^\mu v_1\,q^\nu 
- 4\bar{v}_2(\gamma\inn D \inn q ) v_1\, g^{\mu \nu} 
\right] (\varepsilon \inn D)_{\mu\nu}
\end{eqnarray}

It is straightforward to extend this program to other combinations of 
Neveu-Schwarz and Ramond vertex operators.

It is curious that these amplitudes resemble the Veneziano amplitude,
but occupy an intermediate position between
the conventional three-point and four-point amplitudes 
as far as the number of
kinematic invariants goes. This is a consequence of the
``partial conservation of momentum'' unique to D-branes.

\begin{figure}[t]
\centerline{\psfig{figure=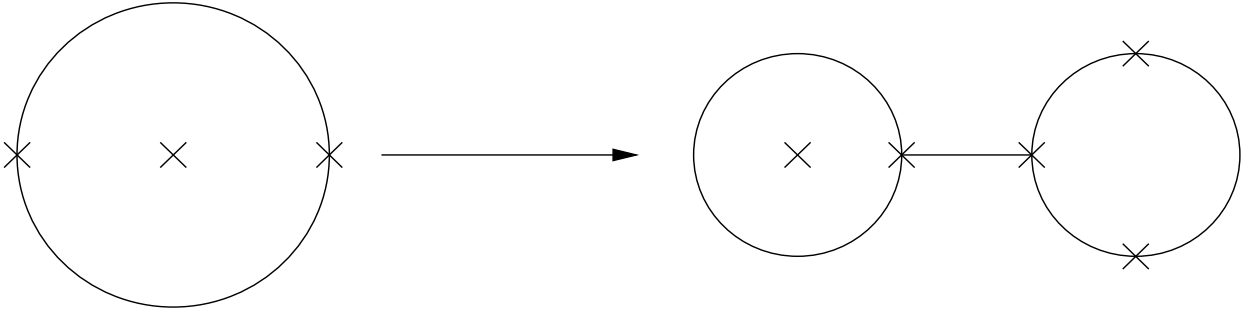,width=\hsize}}
\caption{Factorization of the world sheet giving rise to poles in
$t$.}
\label{figa}
\end{figure}

There are other interesting new features of these amplitudes.  They
decay exponentially for large $t$, which
indicates the softness of
strings at high energies, and have a sequence of poles at half integer
values of $t$. What is surprising is that they also contain a sequence
of zeros for integer values of $t$.  The special role played by these
values of $t$ can be attributed to the massive open string states
appearing in the operator product expansion of the open string vertex
operators when they collide on the world sheet (see figure
(\ref{figa})).
These states have masses
\begin{equation}
m^2 = n/\alpha' = n/2
\end{equation}
for integer $n$, and indeed
\begin{equation}
t = -2 p_1 \inn p_2 = -(p_1 + p_2)^2 = m^2 =  n/2.
\end{equation}
Interestingly, for even $n$ these amplitudes have zeros instead of
poles, indicating that these states do not propagate in the internal
line of figure 1.
Thus, due to the special kinematics of this process, we find
an interesting $Z_2$ 
selection rule. This interplay between zeros and poles is intimately
connected with the exponential decay of the amplitude at large $t$.

\section{Discussion}

In this paper we studied leading order interactions involving the
massless modes of D-branes, the world volume gauge fields and the
scalar fields describing transverse oscillations.  We found that their
four-point functions are identical to those found in type I
theory. They exhibit infinite series of poles due to the massive
D-brane excitations. This implies that, at momenta comparable to
$1/\sqrt{\alpha'}$, the world volume dynamics of D-branes departs
considerably from that dictated by the Dirac-Born-Infeld effective
action.

Our most interesting result concerns the coupling of excited
D-branes to the outside world. The leading such effect is the
three-point function for two massless internal modes and one massless
closed string, which depends on the total longitudinal
momentum-squared, $t$.  For low $t$ the results agree with the
D-brane effective action. Consider, for instance, the coupling of two
transverse oscillations to a graviton. For $t\ll 1$, the exact 
three-point function,
\begin{equation}
A \sim 
\frac{\Gamma(1-2t)}{\Gamma(1 -t)^2}\,
t\, \varepsilon_{ij} \xi_1^i \xi_2^j \rightarrow
t\, \varepsilon_{ij} \xi_1^i \xi_2^j
\end{equation}
This low $t$ approximation coincides with the amplitude obtained
from the following term in the effective action,
\begin{equation}
\int d^{p+1} x \partial \phi^i \partial \phi^j G_{ij}
\end{equation}
In \cite{cm} this low $t$ behavior was found to be crucial for obtaining
the correct properties of the outgoing Hawking radiation.

The exact three-point function
exhibits an infinite sequence of zeros and poles for $t$ comparable to
$1/\sqrt{\alpha'}$ (our convention is to set $\alpha'=2$).  If we
average over the zeros and poles, which is usually done by giving $t$
a moderate imaginary part, then for large $t$ the coupling falls off
exponentially. This is a purely string theoretic effect, which is
physically due to increasing effective thickness of D-branes at high
$t$.\footnote{A related phenomenon, the Regge behavior of high energy
scattering from D-branes, has been observed in previous calculations
\cite{KT,GHKM,Bachas}. On the other hand, there is evidence that at
very low energies the effective size of D-branes becomes very small,
much smaller than the string scale
\cite{Bachas,Shenker,DanFerrSund,PouliotKabat}.}  The fact that highly
energetic internal modes are only weakly coupled to the outside world
is, nevertheless, quite surprising.  Perhaps this gives us a first
indication as to why highly non-extremal black holes are very long
lived in string theory. The entropy of non-extremal black holes has
been a subject of active research lately
\cite{cm,ghas,gkp,juan,BLMPSV96}.  D-brane calculations of their decay
are an interesting new tool, and we hope to return to more detailed
calculations in the near future.

\section*{Acknowledgements}

We are grateful to V.~Balasubramanian, C.~Callan, S.~Gubser,
J.~Maldacena, and V.~Periwal for illuminating discussions.  This work
was supported in part by DOE grant DE-FG02-91ER40671, the NSF
Presidential Young Investigator Award PHY-9157482, and the James
S. McDonnell Foundation grant No.  91-48.

\end{document}